\documentclass[a4paper]{article}

\usepackage{INTERSPEECH2021}
\usepackage{url}
\usepackage{amsmath,graphicx}

\usepackage{xcolor}
\usepackage{multirow}
\usepackage{float}
\usepackage{cite}
\usepackage{amssymb}
\usepackage{algorithm}  
\usepackage{algpseudocode}  
\usepackage{microtype}
\usepackage{paralist}
\usepackage{hyperref}

\title{VQ-T: RNN Transducers using Vector-Quantized Prediction Network States}
\name{Jiatong Shi$^{1}$\thanks{Jiatong Shi performed the work during his internship at IBM Research AI.}, George Saon$^{2}$, David Haws$^{2}$, Shinji Watanabe$^{1}$, Brian Kingsbury$^{2}$}
\address{
    $^{1}$ Language Technologies Institute, Carnegie Mellon University, PA, USA\\
    $^{2}$ IBM Research AI, Yorktown Heights, NY, USA 
} 
\email{jiatongs@cs.cmu.edu, \{gsaon, dhaws, bedk\}@us.ibm.com}

\begin{document}

\maketitle
\begin{abstract}
Beam search, which is the dominant ASR decoding algorithm for end-to-end models, generates tree-structured hypotheses. However,  recent studies have shown that decoding with hypothesis merging can achieve a more efficient search with comparable or better performance. But, the full context in recurrent networks is not compatible with hypothesis merging. We propose to use vector-quantized long short-term memory units (VQ-LSTM) in the prediction network of RNN transducers. By training the discrete representation jointly with the ASR network, hypotheses can be actively merged for lattice generation. Our experiments on the Switchboard corpus show that the proposed VQ RNN transducers improve ASR performance over transducers with regular prediction networks while also producing denser lattices with a very low oracle word error rate (WER) for the same beam size. Additional language model rescoring experiments also demonstrate the effectiveness of the proposed lattice generation scheme.
\end{abstract}
\noindent\textbf{Index Terms}: Lattice Generation; RNN Transducer; Vector-quantized LSTM

\section{Introduction}
\label{sec: intro}

End-to-end automatic speech recognition (ASR) has been a hot research direction. An end-to-end system directly maps acoustic features to linguistic units, reducing the workload of combining different knowledge sources. At the same time, it has shown equivalent or better performance over conventional hidden Markov model (HMM) based approaches \cite{chiu2018state, pham2019very, guo2021recent}. According to recent literature,  we have observed two mainstream modeling directions for end-to-end ASR: attention-based encoder-decoder (AED) \cite{chorowski2015attention, chan2016listen, watanabe2017hybrid}, and RNN transducer (RNN-T) \cite{graves2012sequence, rao2017exploring, saon2021advancing}.

Beam search has been the main decoding strategy for both HMM-based approaches and end-to-end systems. Using a limited beam size $\mathcal{H}$, the search generates an $N$-best list with a computational cost of $\mathcal{O}(\mathcal{H})$. However, $N$-best lists are insufficient for some downstream applications or post-processing stages, such as
\begin{inparaenum}[(1)]
\item language model rescoring~\cite{ljolje1999efficient, sak2010fly, rybach2017lattice};
\item downstream processing of ASR output (e.g., translation\cite{kumar2014some} and keyword spotting \cite{kingsbury2013high, rosenberg2017end});
\item confusion network generation~\cite{mangu2000finding, hakkani2006beyond}; and 
\item sequence discriminative training \cite{normandin1996maximum, povey2002, gibson2006hypothesis}.
\end{inparaenum}

Hypothesis merging (or path recombination) is the main difference between regular beam search and lattice-based search. For conventional methods, Rybach et al. divided merging procedures into two categories \cite{rybach2017lattice}: the phone-pair approach \cite{ljolje1999efficient, mohri2008speech} and the $N$-best history approach \cite{saon2005anatomy, chen2006advances, soltau2009dynamic}. However, neither are suitable for end-to-end models because they require additional n-gram language models and HMM-based acoustic models to generate lattices. In order to generate lattices for end-to-end ASR models, recent works proposed approaches that restrict the hypothesis merging context \cite{zapotoczny19, prabhavalkar2021less} with convolutional and recurrent decoders. Zapotoczny et al. \cite{zapotoczny19} replaced the recurrent decoder with a temporal convolutional network (TCN) in AED networks. Since there is a fixed-length context, this method can merge hypotheses as efficiently as with an n-gram language model. Their experiments on the WSJ dataset indicate that character-based lattice decoding can consistently outperform beam decoding in terms of WER for a similar beam size. Prabhavalkar et al. \cite{prabhavalkar2021less} further investigated hypothesis merging for recurrent neural network transducer (RNN-T) with a full-context prediction network. They approximately merge hypotheses with the same last few byte pair encoding (BPE) labels, showing improvements in both 1-best and oracle WER. 

This work proposes the use of vector-quantized long short-term memory units (VQ-LSTM) to generate lattices for RNN-Ts. The VQ layers are added to the LSTM cells and are jointly optimized with the transducer objective. We show that, with VQ-LSTM, we can actively merge hypotheses for lattice generation while conditioning on the full context. We investigate both systems in \cite{zapotoczny19, prabhavalkar2021less}, and provide a comparison for different types of encoders (LSTM and conformer) and linguistic units (characters and BPE units). Experiments on the Switchboard 300 hours corpus show that our method improves ASR performance on multiple test sets despite using compressed context information. The method also results in denser lattices with a significantly lower oracle WER. The generated lattices from VQ-based systems show improvements for language model rescoring as a downstream task.

\section{RNN Transducer Formulation}
\label{sec: formulation}
In RNN-T, the objective function to be minimized is the negative log-likelihood (NLL) loss, defined as $-\log p(\vec{y}|\vec{x})$ \cite{graves2012sequence, bagby2018efficient}. $\vec{x}=(x_1, ..., x_T)$ is an input speech feature sequence with length $T$,  and $\vec{y}=(y_0, y_1, ..., y_U)$ is an output label sequence with length $U+1$, where $y_u \in \mathcal{Y}$ with label vocabulary $\mathcal{Y}$ and $y_0$ is the start token of a label sentence. Symbol prediction probability $p(\vec{y}|\vec{x})$ can be elaborated into the summation over all possible alignments $\vec{a}$:
\begin{equation}
    p(\vec{y}|\vec{x}) = \sum_{\vec{a} \in \mathcal{B}^{-1}(\vec{y})} p(\vec{a}|\vec{x})
\end{equation}
where $\vec{a} \in \mathcal{Y} \cup \{\phi\}$ is a sequence of symbols augmented by \textit{blank} symbols $\phi$, and function $\mathcal{B}$ maps alignment $\vec{a}$ to the output sequence $\vec{y}$. The speech features $\vec{x}$ are encoded into hidden states $\vec{m} = (m_1, m_2, ..., m_T)$ by an \textit{encoder} network. The partial output sequence $\vec{y}_{u} = (y_0, y_1, ..., y_{u-1})$ is encoded into hidden states $\vec{g}_{u} = (g_1, g_2, ..., g_{u})$ by  a \textit{prediction network}. 

$p(\vec{a}|\vec{x})$ is given by a \textit{joint network} based on the following factorization:
\begin{equation}
\label{eq: factor}
\begin{split}
    p(\vec{a}|\vec{x}) = p(\vec{a}|\vec{m}) \triangleq \prod_{i=1}^{T+U} p(a_i|m_{t_i}, \mathcal{B}(a_1, ..., a_{i-1})) = \\
    = \prod_{i=1}^{T+U}p(a_i|m_{t_i}, y_0, ..., y_{u_{i-1}}) = \prod_{i=1}^{T+U} p(a_i|m_{t_i}, g_{u_i}),
\end{split}
\end{equation}
where $t_i$ and $u_i$ are corresponding $t$ and $u$ at the alignment step $i$. Based on the factorization, the NLL loss can be efficiently computed with the forward-backward algorithm, which has computational complexity bounded by $\mathcal{O}(TU)$. 


\section{VQ-LSTM and Lattice Generation}


\begin{figure}[tbp]
    \centering
    \centerline{\includegraphics[width=8.5cm]{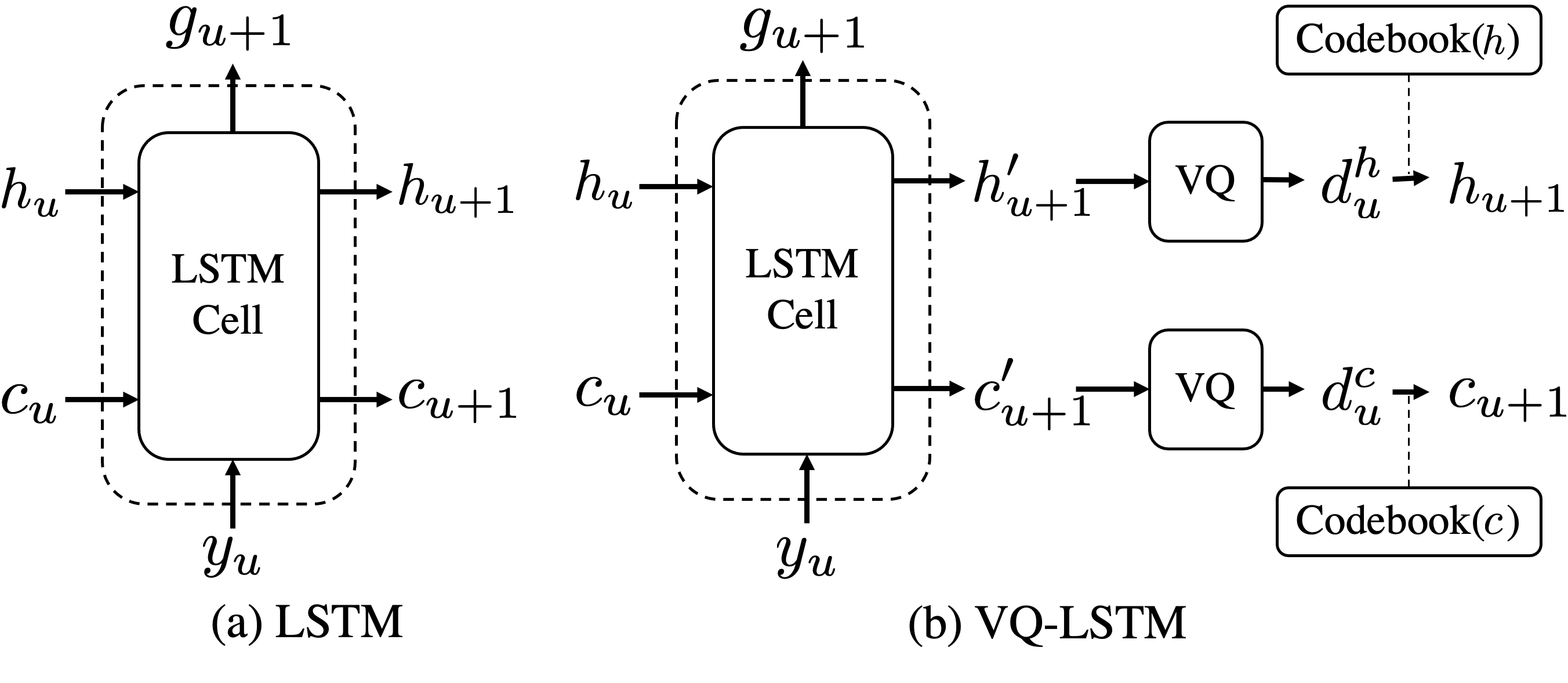}}
\vspace{-15pt}
\caption{Structure comparison between LSTM and VQ-LSTM.}
\label{fig:comparison}
\vspace*{-5mm}
\end{figure}

\subsection{Very-limited-context (VLC) Model}
\label{ssec: VLC model}
As noted in Section~\ref{sec: intro}, lattice generation in end-to-end ASR models relies on hypothesis merging. Compared to the RNN-T with full-context LSTM layers, several recent studies have found that the \textit{prediction network} can achieve comparable performances with limited context or simple structures \cite{prabhavalkar2021less, zapotoczny19, variani2020hybrid, botros21_interspeech}. 
Zapotoczny et al. \cite{zapotoczny19} proposed to apply TCN for lattice generation in the attention-based encoder-decoder (AED) architecture so as to limit the context received in the decoder network. Following this direction, we employ simple convolutional layers in RNN-T \textit{prediction network}s. 
We find a left context of only two tokens could already reach reasonable ASR performance, so we introduce this model as a reference model for our following discussion, namely as a very-limited-context (VLC) model.

\subsection{VQ-LSTM}
\label{ssec: vq-lstm}

Hypothesis merging for lattice generation is straightforward in HMM-based hybrid systems thanks to the limited context in n-gram language models. However, the hidden state of LSTM \textit{prediction network} for further expansion is a tuple of continuous vectors $h_u$ and $c_u$ at iteration $u$. A compromise is to train the LSTM-based \textit{prediction network} only on a limited context, which has been shown to degrade performance \cite{prabhavalkar2021less}. As an alternative in this work, we apply vector quantization to the LSTM hidden and cell states.

Vector quantization (VQ) has been successfully applied in several areas, especially representation learning \cite{baevski2019vq, chorowski2019unsupervised, dunbar2020zero, robine2020smaller}, but more recently as a useful inductive bias~\cite{liu2021dvnc} to enhance generalization. In most cases, the VQ process acts as a feature extractor, encoding intermediate hidden states from the network into a more compact discrete representation. Previous literature explored distilling the probabilistic model into a weighted finite state transducer (wFST) so as to perform rescoring for ASR and other natural language processing tasks \cite{lecorve12b_interspeech, suresh2021approximating}. However, the performance of these methods is likely to be bounded by the performance of the original probabilistic model, given that the distillation was not optimized with the same objective as the probabilistic model.

In this work, we propose to apply a trainable VQ layer to the LSTM states to enable lattice generation without performance degradation for 1-best decoding. As illustrated in Fig.~\ref{fig:comparison}(b), we denote $h_u, c_u, g_u \in \mathbb{R}^{D}$, where $D$ is the dimension of corresponding vectors. Then, an iteration of original LSTM is defined as:
\begin{equation}
    g_{u+1}, (h_{u+1}, c_{u+1}) = \mathrm{LSTM}(y_u, h_u, c_u).
\end{equation}
As shown in Fig.~\ref{fig:comparison}(b), to apply vector quantization to LSTM, we add two vector quantizers, resulting in a recurrent network with discrete hidden and cell states $d_u^h \in \mathbb{N}, d_u^c \in \mathbb{N}$. With the quantizers, an iteration $u$ of VQ-LSTM is defined as:
\begin{align}
  g_{u+1}, (h'_{u+1}, c'_{u+1}) & = \mathrm{LSTM}(y_u, h_u, c_u) \label{eq: lstm_vq},\\
 (d^{h}_{u}, d^{c}_{u}) & = \mathrm{VQ}(h'_{u+1}, c'_{u+1}) \label{eq: lstm_vq-vq}\\
 (h_{u+1}, c_{u+1}) & = \mathrm{Codebook}(d^h_u, d^c_u). , \label{eq:codebook} 
\end{align}
The detailed implementation of $\mathrm{VQ}(\cdot)$ in Eq.~\eqref{eq: lstm_vq-vq} follows VQ-wav2vec \cite{baevski2019vq}: we first employ a few fully-connected layers to project the states $h'_{u+1}, c'_{u+1}$ into logits. Next, we apply GumbelSoftmax to convert the input vector into probability and use argmax to select the highest probability index as $d^h_u, d^c_u$ in the related codebook. The corresponding vector obtained by $\mathrm{Codebook}(\cdot)$ in Eq.~\eqref{eq:codebook} is then passed to the joint network. 
The discrete representations $d^h_{u},d^c_{u}$ are used for the hypothesis merging step in lattice generation.

\begin{figure*}[tbp]
    \centering
    \centerline{\includegraphics[width=14cm]{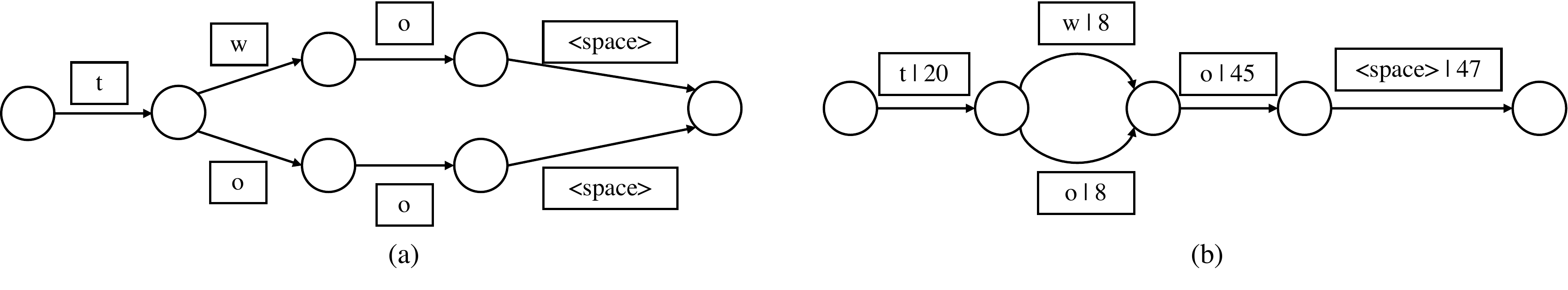}}
\vspace{-15pt}
\caption{Lattice examples: (a) Very-limited-context (VLC) model from Sec~\ref{ssec: VLC model}. (b) VQ-based model from Sec~\ref{ssec: vq-lstm} (numbers represent codebook indices). As shown in (b), VQ-based model can produce the same discrete code (i.e., $d^h_u, d^c_u$), even when contexts are different.}
\label{fig:lattice example}
\vspace{-15pt}
\end{figure*}

\subsection{Lattice Generation Algorithm}

The VQ-based lattice generation is based on alignment length synchronous decoding (ALSD) with hypothesis merging \cite{saon2020alignment}. Its pseudo code is shown in Algorithm~\ref{alg:lattice-gen}.

The lattice $\mathcal{L}$ is initialized as an empty finite state acceptor (FSA). A hypothesis is defined by the tuple $(n, \vec{y}_u, (h_u, c_u), \delta_{t,u}(\vec{y}_u))$, consisting of its last node $n$ in the lattice, partial output sequence $\vec{y}_u$, \textit{prediction network} states $(h_u, c_u)$, and hypothesis score $\delta_{t,u}(\vec{y}_u)$. The score $\delta_{t,u}(\vec{y}_u)$ stands for the posterior probability of sequence $\vec{y}_u$ at time $t$ and position $u$. As in \cite{saon2020alignment}, we define two hypothesis sets $A$ and $B$ to store beams of hypotheses for alignment steps $i$ and $i-1$, respectively. Hypothesis set $F$ is also introduced to store final hypotheses. The set $B$ starts with the initial hypothesis $(n_{\text{start}}, \phi, (h_{0}, c_{0}), 1)$, while $A$ and $F$ start from an empty set.

\begin{algorithm}[htb]  
  \caption{ VQ-based Lattice Generation}  
  \label{alg:lattice-gen}  
  \begin{algorithmic}[1]  

    \Require  
    
      Encoder states: $\vec{m}$;  
      
      Beam size:  $\mathcal{H}$; 
      
      Max output sequence length: $U_{max}$;
       
    \Ensure  
      Hypotheses sorted by likelihood: $hyps$; 
      
      Lattice: $\mathcal{L}$;
      
    \State $\mathcal{L}, n_{\text{start}}, n_{\text{end}}=$FSA()
    \State $B=\{(n_{\text{start}}, \phi, (h_{0}, c_{0}), 1) \}$; $F = \{\}$; $A = \{\}$
    
    \For{$i=1...T + U_{max}$}
        \State $A=$ExpandHypothesis$(B, \vec{m})$
        \State $A,F =$MergeVQState$(A, \mathcal{H})$
        \State $B =$Prune$(A)$
        \State $\mathcal{L}, B=$UpdateLattice$(B, \mathcal{L})$

    \EndFor
    
    \State \Return sorted($F$) and $\mathcal{L}$; 
  \end{algorithmic}  

\end{algorithm}

During the main loop, the alignment step $i$ is iterated from $1$ to $T + U_{max}$, where $U_{max}$ is an estimate of the maximum output sequence length.  The merge actions are recorded in lattice $\mathcal{L}$. Four major functions in Algorithm~\ref{alg:lattice-gen} include:

\noindent
\textbf{(1)~ExpandHypothesis} (line 4): each hypothesis is expanded over the vocabulary size $|\mathcal{Y}|$. Compared to the algorithm in \cite{saon2020alignment}, for all potential hypotheses, we apply additional \textit{prediction network} forward-passes (i.e., Eq.~\eqref{eq: lstm_vq} and Eq.~\eqref{eq: lstm_vq-vq}) to extract discrete context representations $(d^h_{u}, d^c_{u})$ discussed in Sec.~\ref{ssec: vq-lstm}.

\noindent
\textbf{(2)~MergeVQState} (line 5): hypothesis merging with the same state (i.e., state code $(d^h_{u}, d^c_{u})$ for VQ-based system). During the merge, the posterior probability is summed over merging hypotheses, and the hypothesis with the maximum probability are selected (i.e., keep its corresponding $(d^h_{u}, d^c_{u})$ for the merged hypothesis) for further expansion.

\noindent
\textbf{(3)~Prune} (line 6): pruning with respect to beam size. 

\noindent
\textbf{(4)~UpdateLattice} (line 7): updating lattice. For merged hypotheses, the additional arcs have the same target node, while for others, the arcs link to different target nodes. 

The VLC model has similar lattice generation procedures, but it does not record $(d^h_{u}, d^c_{u})$ over step $u$, nor does it perform the additional forward-pass in ``ExpandHypothesis''.

Examples of generated lattices are shown in Figure~\ref{fig:lattice example}. VLC needs to have same limited context (i.e., context of ``o'' ``space'' for hypothesis merging. The VQ-based model does not have this requirement and have the possibility of generating the same discrete hidden state (i.e., ``8'' for ``w'' and ``o'' in Fig.~\ref{fig:lattice example}) for hypotheses merging.

\section{Experiments}

\begin{table*}
\centering
\begin{tabular}{lcccccc}
\toprule
\multirow{2}{*}{\textbf{Model}} & \multirow{2}{*}{\textbf{StateNum}} & \multirow{2}{*}{\textbf{Encoder}} & \multirow{2}{*}{\textbf{Outputs}} & \multicolumn{3}{c}{\textbf{Evaluation (WER)}}  \\ 
& & & & \textbf{Hub5'00} & \textbf{Hub5'01} & \textbf{RT'03}\\
\midrule
\multirow{4}{*}{\textbf{Baseline}} &  \multirow{4}{*}{$\infty$} & \multirow{2}{*}{LSTM} & Char & 11.8 & 12.1 & 15.3\\
 & &  & BPE & 12.1 & 12.3 & 15.2\\
 &  & \multirow{2}{*}{Conformer} & Char & 11.4 & 11.2& 14.4 \\
 &  &  & BPE & 11.3 & \textbf{10.7} & 14.9 \\
 \midrule
 \multirow{4}{*}{\textbf{VLC}} &  \multirow{4}{*}{$|\mathcal{Y}|^2$} &  \multirow{2}{*}{LSTM} & Char & 11.7 & 12.2 & 15.1 \\
 & &  & BPE & 11.4 & 12.1 & 14.9 \\
 &  & \multirow{2}{*}{Conformer} & Char & 11.5 & 11.4 & 14.6\\
 &  &  & BPE & 11.4 & 10.9 & 14.5\\
 \midrule
 \multirow{4}{*}{\textbf{VQ-T}} &  \multirow{4}{*}{640$^4$} & \multirow{2}{*}{LSTM} & Char & 11.7 & 11.7 & 14.8 \\
 & &  & BPE & 12.0 & 12.1 & 15.1\\
 &  & \multirow{2}{*}{Conformer} & Char & 11.3 & 10.9 & 14.2 \\
 &  &  & BPE & \textbf{11.1} & 10.9 & \textbf{13.6} \\

\bottomrule
\end{tabular}

\caption{\label{tab: WER} Word error rate (WER) at beam size 16 for different models: StateNum is the total number of states in \textit{prediction network}. $|\mathcal{Y}|$ for VLC is the vocabulary size, which is 45 for characters and 681 for BPE units.}
\vspace{-20pt}
\end{table*}



\begin{table}
\centering
\begin{tabular}{l|cc}
\toprule Model & Lattice Density  \\ 
\midrule
Baseline         & 31.5 \\
VLC              & 25.4 \\
VQ-T             & 165.5   \\
\bottomrule
\end{tabular}

\caption{\label{tab: Lattice Density} Lattice density at beam size 8 for models with LSTM encoders over characters.}
\vspace{-20pt}
\end{table}


\subsection{Experimental Settings}
\label{ssec: experimental setings}

\textbf{Datasets}: Experiments are conducted on the Switchboard speech corpus, which has 300 hours of English telephone conversations between two speakers on a preassigned topic. We use the same data segmentation and transcript preparation as the Kaldi \texttt{s5c} recipe~\cite{povey2011kaldi}. For the evaluation set, we use the Hub5'00, Hub5'01, and RT'03 test sets with Kaldi-style measurement of the WER.

\noindent
\textbf{Features and data augmentation}: We keep the same feature extraction setup and augmentation strategies as in~\cite{saon2021advancing}. To be specific, features are derived from log-Mel filter-bank features and i-vectors. Speed perturbation and SpecAugment are also applied during the training. We evaluate our models for both character and byte-pair encoding (BPE). The BPE units are extracted using 100-step byte-pair encoding.\footnote{https://github.com/rsennrich/subword-nmt} The vocabulary sizes $|\mathcal{Y}|$ for character and BPE models are 45 and 681, respectively.

\noindent
\textbf{Model architectures}: Three models are used in the experiments: \textbf{Baseline}, \textbf{VLC}, and \textbf{VQ-T}. The \textbf{Baseline} model is an RNN-T which uses an \textit{encoder} based on a stack of 6 bidirectional LSTMs with 640 units per layer per direction initialized with a model trained with connectionist temporal classification loss \cite{audhkhasi2019forget}. The \textit{prediction network} is a single-layer unidirectional LSTM with 768 units (i.e., $D = 768$). The \textit{joint network} takes the input from the \textit{encoder} and \textit{prediction network}s and projects them into 256-dimensional vectors. The projected vectors are multiplied together element-wise~\cite{saon2021advancing}, passed through a $\tanh$ activation, and projected to the vocabulary $\mathcal{Y}$ and \textit{blank} symbol $\phi$. The \textbf{Baseline} with conformer encoders has the same \textit{prediction network} configuration as \textbf{Baseline}, but the \textit{encoder} has 12 layers, a hidden dimension of 384, a convolutional kernel of size 31, and six 96-dimensional attention heads per layer.

The \textbf{VLC} model adopts the same \textit{encoder} architecture as \textbf{Baseline}, but with a convolutional \textit{prediction network}. The \textit{prediction network} in \textbf{VLC} has two parallel 256-dimensional-convolutional layers, one with a $\tanh()$ nonlinearity and one bias-free skip connection. Both have a two-linguistic-unit left context. The output of the \textit{prediction network} is the sum of the two convolutional layers.

We perform vector quantization via GumbelSoftmax as introduced in \cite{jang2016categorical}. Three hyper-parameters are introduced for the VQ layers: $vq\_group$ is the number of groups in each VQ layer; $vq\_vars$ is the number of codebook vectors per group; $vq\_depth$ is the number of fully connected layers before converting the input vector into logits. These three hyper-parameters have the strongest influence in our experiments, so other hyper-parameters such as training temperature and GumbelSoftmax coefficients are set to default settings from VQ-wav2vec.
\textbf{VQ-T} follows the same architecture as \textbf{Baseline}, but with additional VQ layers in LSTM-based \textit{prediction network}s. The default setting for VQ layer is with $vq\_group$ of 2, $vq\_vars$ of 640, $vq\_depth$ of 1. With these defaults, the total number of additional parameters introduced by the VQ-layer is less than 1M, less than 5\% of the total model parameters.

\noindent
\textbf{Training}: All models are trained on 2 V100 GPUs with minibatches of 32 utterances per GPU, the AdamW optimizer, and the OneCycle scheduler. VQ models are trained for 30 epochs, while the other models are trained for 20.\footnote{Our empirical studies show that VQ layers need more epochs to converge.} The peak learning rate is 0.0005.

\noindent
\textbf{Decoding}: In \cite{saon2020alignment}, the decoding algorithms for RNN-Ts are categorized into two types: time synchronous decoding (TSD) and ALSD. Experiments in \cite{saon2020alignment} indicate that ALSD has similar accuracy as TSD, but is much faster. Therefore, we choose ALSD as our base decoding strategy.


\noindent
\textbf{Lattice generation}: Both \textbf{VLC} and \textbf{VQ-T} models generate lattices by merging hypotheses with the same discrete hidden states, in contrast to \cite{prabhavalkar2021less} where full context \textit{prediction networks} are used but hypotheses are merged based on the same limited context from running hypotheses. For comparison, we also generate lattices with \textbf{Baseline} by fixing the same context as \textbf{VLC} for merging. Since the lattice generation for \textbf{VQ-T} needs to conduct one extra forward pass for each decoding step, we only conduct experiments on character-based lattices for efficiency.

\noindent
\textbf{Language model rescoring}: To evaluate the lattice quality, we also conduct language model rescoring over lattices. We apply a three-layer 512 dimensional LSTM language model trained on 25M words of the Fisher and Switchboard transcripts. We prune the \textbf{VQ-T} lattices before rescoring with a relative margin\footnote{10\% over the unit with the least log probability.} for arcs that share the same source and target node.

\noindent
\textbf{Evaluation metric}: Besides WER for ASR performance, we adopt two metrics for lattice evaluation: lattice density and oracle WER. Lattice density is the average number of lattice arcs per frame. Oracle WER has been used to evaluate lattice quality by finding the lowest word error hypothesis in the lattice.

\begin{figure}[t]
    \centering
    \centerline{\includegraphics[trim=0 10 0 0,clip,width=7cm]{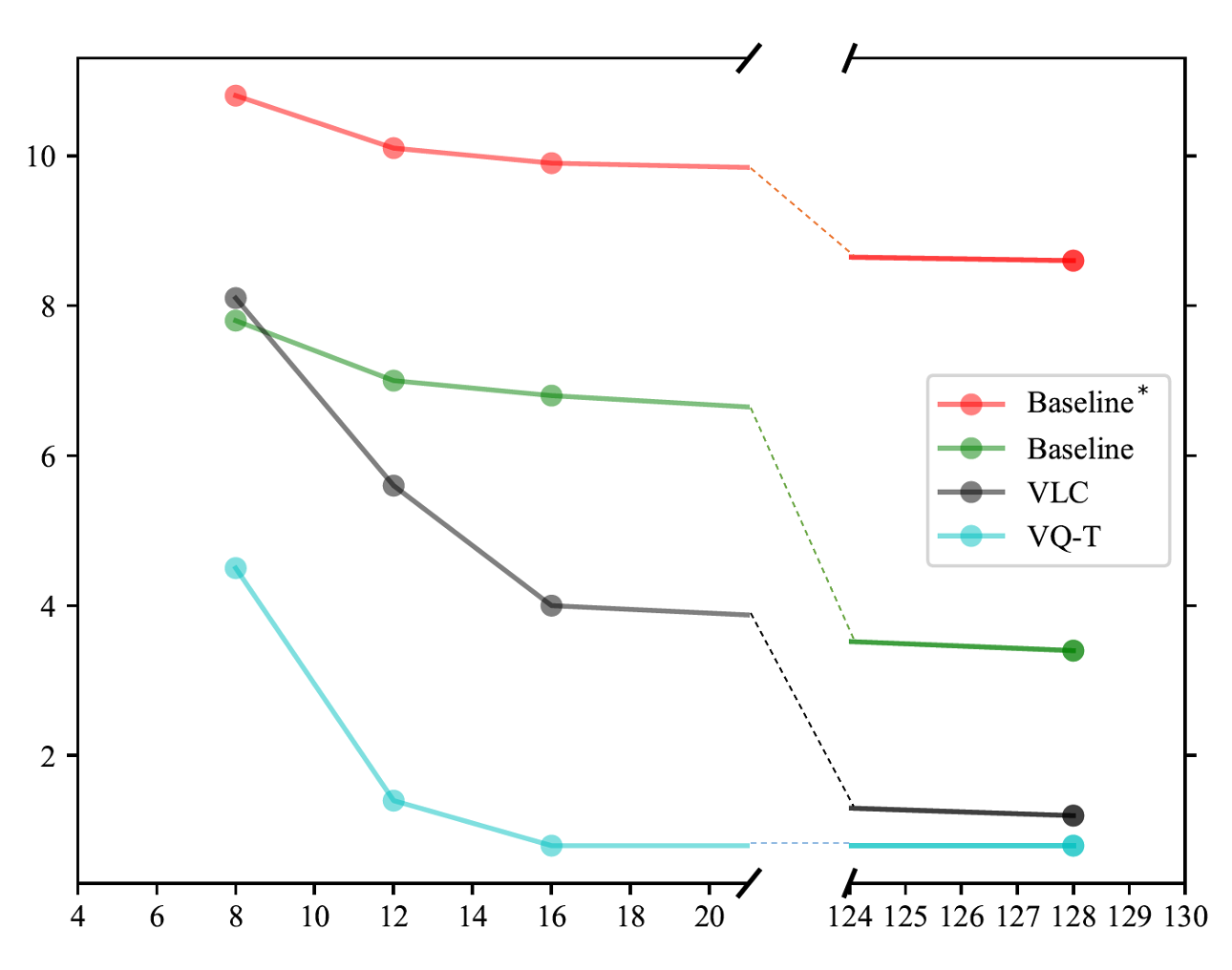}}
    \vspace{-10pt}
\caption{Oracle WER for models with different beam sizes: the models are with LSTM encoders over characters. Baseline$^*$ is based on tree structure from beam search.}
\label{fig:oracle wer}
\vspace{-20pt}
\end{figure}

\begin{table}[t]
\centering
\begin{tabular}{l|cc}
\toprule Model & LSTM & Conformer  \\ 
\midrule
Baseline         & 11.3(0.5) & 11.0(0.4) \\
VLC              & 11.2(0.5) & 10.9(0.6) \\
VQ-T             & \textbf{11.0(0.7)} & \textbf{10.8(0.5)} \\
\bottomrule
\end{tabular}
\caption{\label{tab: lm rescoring} Language model rescoring for character-based models on Hub5'00. Absolute improvements from rescoring are shown in parentheses.}
\vspace{-30pt}
\end{table}

\subsection{Results and Discussion}

The WER for different models is shown in Table~\ref{tab: WER}. The \textbf{VQ-T} model with conformer encoders over BPE units reaches the overall best performance of 11.1 and 13.6 for Hub5'00 and RT'03. We also observe similar performances on Hub5'01 to the baseline with full context. Aligning with findings in \cite{variani2020hybrid, botros21_interspeech}, \textbf{VLC} model performs comparably to the \textbf{Baseline} model. At the same time, the learnable \textbf{VQ-T} model performs slightly better in most cases.

Table~\ref{tab: Lattice Density} shows the lattice density for three sample models with LSTM encoders over characters. Compared to other models, our proposed \textbf{VQ-T} models have significantly denser lattices. The major reason is that \textbf{VQ-T} performs more active hypothesis merging for similar context. Fig.\ref{fig:oracle wer} illustrates the oracle WER from lattices generated from the three models over different beam sizes. The results are similar to Table~\ref{tab: Lattice Density}. Because of the denser lattices, the \textbf{VQ-T} oracle WERs converge considerably faster when increasing the beam size. We also observe much lower oracle WERs at convergence for \textbf{VQ-T} systems.


Table~\ref{tab: lm rescoring} shows the WER from lattice rescoring. While rescoring leads to better transcription accuracy in all cases, the best performance is obtained with the \textbf{VQ-T} models.


\section{Conclusion and Future Work}
In this work, we use vector-quantization of prediction network states to simplify lattice generation with RNN-Ts. The method efficiently compresses full context into discrete states, providing natural merge points for hypothesis merging in lattice generation. Our experiments on the Switchboard conversational corpus show that the proposed VQ-based transducer improves the performance of beam search 1-best decoding and generates dense, efficient lattices with a very low oracle WER for small beam sizes. We also verify the lattice quality using language model rescoring as a downstream task. One major challenge with the proposed method, however, is the increased decoding time, given that we introduce extra forward steps in the lattice generation algorithm. Since the forward pass needs to iterate over the entire vocabulary set, it is diffcult to extend the lattice generation to BPE units or words. Therefore, our future work will focus on improving decoding and VQ-based lattice generation efficiency with BPE units. Another direction is to use the lattices for more downstream tasks (e.g., translation, keyword spotting, and confusion network processing).

\bibliographystyle{IEEEtran}

\bibliography{mybib}

\begin{thebibliography}{10}
\providecommand{\url}[1]{#1}
\csname url@samestyle\endcsname
\providecommand{\newblock}{\relax}
\providecommand{\bibinfo}[2]{#2}
\providecommand{\BIBentrySTDinterwordspacing}{\spaceskip=0pt\relax}
\providecommand{\BIBentryALTinterwordstretchfactor}{4}
\providecommand{\BIBentryALTinterwordspacing}{\spaceskip=\fontdimen2\font plus
\BIBentryALTinterwordstretchfactor\fontdimen3\font minus
  \fontdimen4\font\relax}
\providecommand{\BIBforeignlanguage}[2]{{%
\expandafter\ifx\csname l@#1\endcsname\relax
\typeout{** WARNING: IEEEtran.bst: No hyphenation pattern has been}%
\typeout{** loaded for the language `#1'. Using the pattern for}%
\typeout{** the default language instead.}%
\else
\language=\csname l@#1\endcsname
\fi
#2}}
\providecommand{\BIBdecl}{\relax}
\BIBdecl

\bibitem{chiu2018state}
C.-C. Chiu, T.~N. Sainath, Y.~Wu, R.~Prabhavalkar \emph{et~al.},
  ``State-of-the-art speech recognition with sequence-to-sequence models,'' in
  \emph{ICASSP}, 2018, pp. 4774--4778.

\bibitem{pham2019very}
N.-Q. Pham, T.-S. Nguyen, J.~Niehues, M.~M{\"u}ller \emph{et~al.}, ``Very deep
  self-attention networks for end-to-end speech recognition,''
  \emph{Interspeech}, pp. 66--70, 2019.

\bibitem{guo2021recent}
P.~Guo, F.~Boyer, X.~Chang, T.~Hayashi \emph{et~al.}, ``Recent developments on
  espnet toolkit boosted by conformer,'' in \emph{ICASSP}, 2021, pp.
  5874--5878.

\bibitem{chorowski2015attention}
J.~K. Chorowski, D.~Bahdanau, D.~Serdyuk, K.~Cho \emph{et~al.},
  ``Attention-based models for speech recognition,'' in \emph{NIPS}, 2015, pp.
  577--585.

\bibitem{chan2016listen}
W.~Chan, N.~Jaitly, Q.~Le, and O.~Vinyals, ``Listen, attend and spell: A neural
  network for large vocabulary conversational speech recognition,'' in
  \emph{ICASSP}, 2016, pp. 4960--4964.

\bibitem{watanabe2017hybrid}
S.~Watanabe, T.~Hori, S.~Kim, J.~R. Hershey \emph{et~al.}, ``Hybrid
  ctc/attention architecture for end-to-end speech recognition,'' \emph{IEEE
  Journal of Selected Topics in Signal Processing}, vol.~11, no.~8, pp.
  1240--1253, 2017.

\bibitem{graves2012sequence}
A.~Graves, ``Sequence transduction with recurrent neural networks,''
  \emph{arXiv preprint arXiv:1211.3711}, 2012.

\bibitem{rao2017exploring}
K.~Rao, H.~Sak, and R.~Prabhavalkar, ``Exploring architectures, data and units
  for streaming end-to-end speech recognition with {RNN}-transducer,'' in
  \emph{ASRU}, 2017, pp. 193--199.

\bibitem{saon2021advancing}
G.~Saon, Z.~T{\"u}ske, D.~Bolanos, and B.~Kingsbury, ``Advancing {RNN}
  transducer technology for speech recognition,'' in \emph{ICASSP}, 2021, pp.
  5654--5658.

\bibitem{ljolje1999efficient}
A.~Ljolje, F.~Pereira, and M.~Riley, ``Efficient general lattice generation and
  rescoring,'' in \emph{EuroSpeech}, 1999.

\bibitem{sak2010fly}
H.~Sak, M.~Saraclar, and T.~G{\"u}ng{\"o}r, ``On-the-fly lattice rescoring for
  real-time automatic speech recognition,'' in \emph{Interspeech}, 2010.

\bibitem{rybach2017lattice}
D.~Rybach, M.~Riley, and J.~Schalkwyk, ``On lattice generation for large
  vocabulary speech recognition,'' in \emph{ASRU}, 2017, pp. 228--235.

\bibitem{kumar2014some}
G.~Kumar, M.~Post, D.~Povey, and S.~Khudanpur, ``Some insights from translating
  conversational telephone speech,'' in \emph{ICASSP}, 2014, pp. 3231--3235.

\bibitem{kingsbury2013high}
B.~Kingsbury, J.~Cui, X.~Cui, M.~J. Gales \emph{et~al.}, ``A high-performance
  cantonese keyword search system,'' in \emph{ICASSP}, 2013, pp. 8277--8281.

\bibitem{rosenberg2017end}
A.~Rosenberg, K.~Audhkhasi, A.~Sethy, B.~Ramabhadran \emph{et~al.},
  ``End-to-end speech recognition and keyword search on low-resource
  languages,'' in \emph{ICASSP}, 2017, pp. 5280--5284.

\bibitem{mangu2000finding}
L.~Mangu, E.~Brill, and A.~Stolcke, ``Finding consensus in speech recognition:
  word error minimization and other applications of confusion networks,''
  \emph{CSL}, vol.~14, no.~4, pp. 373--400, 2000.

\bibitem{hakkani2006beyond}
D.~Hakkani-T{\"u}r, F.~B{\'e}chet, G.~Riccardi, and G.~Tur, ``Beyond asr
  1-best: Using word confusion networks in spoken language understanding,''
  \emph{CSL}, vol.~20, no.~4, pp. 495--514, 2006.

\bibitem{normandin1996maximum}
Y.~Normandin, ``Maximum mutual information estimation of hidden markov
  models,'' in \emph{ASRU}, 1996, pp. 57--81.

\bibitem{povey2002}
D.~Povey and P.~C. Woodland, ``Minimum phone error and i-smoothing for improved
  discriminative training,'' in \emph{ICASSP}, 2002, pp. 100--105.

\bibitem{gibson2006hypothesis}
M.~Gibson and T.~Hain, ``Hypothesis spaces for minimum bayes risk training in
  large vocabulary speech recognition.'' in \emph{Interspeech}, vol.~6, 2006,
  pp. 2406--2409.

\bibitem{mohri2008speech}
M.~Mohri, F.~Pereira, and M.~Riley, ``Speech recognition with weighted
  finite-state transducers,'' in \emph{Springer handbook of speech processing},
  2008, pp. 559--584.

\bibitem{saon2005anatomy}
G.~Saon, D.~Povey, and G.~Zweig, ``Anatomy of an extremely fast lvcsr
  decoder,'' in \emph{EuroSpeech}, 2005.

\bibitem{chen2006advances}
S.~F. Chen, B.~Kingsbury, L.~Mangu, D.~Povey \emph{et~al.}, ``Advances in
  speech transcription at ibm under the darpa ears program,'' \emph{TASLP},
  vol.~14, no.~5, pp. 1596--1608, 2006.

\bibitem{soltau2009dynamic}
H.~Soltau and G.~Saon, ``Dynamic network decoding revisited,'' in \emph{ASRU},
  2009, pp. 276--281.

\bibitem{zapotoczny19}
M.~Zapotoczny, P.~Pietrzak, A.~Łańcucki, and J.~Chorowski, ``{Lattice
  Generation in Attention-Based Speech Recognition Models},'' in
  \emph{Interspeech}, 2019, pp. 2225--2229.

\bibitem{prabhavalkar2021less}
R.~Prabhavalkar, Y.~He, D.~Rybach, S.~Campbell \emph{et~al.}, ``Less is more:
  Improved {RNN}-t decoding using limited label context and path merging,'' in
  \emph{ICASSP}, 2021, pp. 5659--5663.

\bibitem{bagby2018efficient}
T.~Bagby, K.~Rao, and K.~C. Sim, ``Efficient implementation of recurrent neural
  network transducer in tensorflow,'' in \emph{SLT}, 2018, pp. 506--512.

\bibitem{variani2020hybrid}
E.~Variani, D.~Rybach, C.~Allauzen \emph{et~al.}, ``Hybrid autoregressive
  transducer (hat),'' in \emph{ICASSP}.\hskip 1em plus 0.5em minus 0.4em\relax
  IEEE, 2020, pp. 6139--6143.

\bibitem{botros21_interspeech}
R.~Botros, T.~N. Sainath, R.~David \emph{et~al.}, ``{Tied \& Reduced {RNN}-T
  Decoder},'' in \emph{Proc. Interspeech 2021}, 2021, pp. 4563--4567.

\bibitem{baevski2019vq}
A.~Baevski, S.~Schneider, and M.~Auli, ``vq-wav2vec: Self-supervised learning
  of discrete speech representations,'' \emph{arXiv preprint arXiv:1910.05453},
  2019.

\bibitem{chorowski2019unsupervised}
J.~Chorowski, R.~J. Weiss, S.~Bengio, and A.~van~den Oord, ``Unsupervised
  speech representation learning using wavenet autoencoders,'' \emph{TASLP},
  vol.~27, no.~12, pp. 2041--2053, 2019.

\bibitem{dunbar2020zero}
E.~Dunbar, J.~Karadayi, M.~Bernard, X.-N. Cao \emph{et~al.}, ``The zero
  resource speech challenge 2020: Discovering discrete subword and word
  units,'' in \emph{Interspeech}, 2020.

\bibitem{robine2020smaller}
J.~Robine, T.~Uelwer, and S.~Harmeling, ``Smaller world models for
  reinforcement learning,'' \emph{arXiv preprint arXiv:2010.05767}, 2020.

\bibitem{liu2021dvnc}
D.~Liu, A.~Lamb, K.~Kawaguchi \emph{et~al.}, ``Discrete-valued neural
  communication in structured architectures enhances generalization,'' in
  \emph{Proc. NeurIPS}, 2021.

\bibitem{lecorve12b_interspeech}
G.~Lecorvé and P.~Motlicek, ``{Conversion of recurrent neural network language
  models to weighted finite state transducers for automatic speech
  recognition},'' in \emph{Interspeech}, 2012, pp. 1668--1671.

\bibitem{suresh2021approximating}
A.~T. Suresh, B.~Roark, M.~Riley \emph{et~al.}, ``Approximating probabilistic
  models as weighted finite automata,'' \emph{Computational Linguistics},
  vol.~47, no.~2, pp. 221--254, 2021.

\bibitem{saon2020alignment}
G.~Saon, Z.~T{\"u}ske, and K.~Audhkhasi, ``Alignment-length synchronous
  decoding for {RNN} transducer,'' in \emph{ICASSP}, 2020, pp. 7804--7808.

\bibitem{povey2011kaldi}
D.~Povey, A.~Ghoshal, G.~Boulianne, L.~Burget \emph{et~al.}, ``The kaldi speech
  recognition toolkit,'' in \emph{ASRU}, 2011.

\bibitem{audhkhasi2019forget}
K.~Audhkhasi, G.~Saon, Z.~T{\"u}ske, B.~Kingsbury, and M.~Picheny, ``Forget a
  bit to learn better: Soft forgetting for ctc-based automatic speech
  recognition.'' in \emph{Interspeech}, 2019, pp. 2618--2622.

\bibitem{jang2016categorical}
E.~Jang, S.~Gu, and B.~Poole, ``Categorical reparameterization with
  gumbel-softmax,'' \emph{arXiv preprint arXiv:1611.01144}, 2016.

\end{thebibliography}


\end{document}